\documentclass[prl,twocolumn,amsmath,amssymb,superscriptaddress]{revtex4}

\usepackage{hyperref}

\usepackage{graphicx}
\usepackage{epsfig}
\usepackage{epstopdf}
\usepackage{cleveref}
\usepackage{amsmath}
\usepackage{bm}
\usepackage{amssymb}

\usepackage{import}
\usepackage{color}

\usepackage{float}

\begin{document}

\title{Intrinsic lineshape of the Raman 2D-mode in freestanding graphene monolayers}

\author{St\'ephane Berciaud}
\email{stephane.berciaud@ipcms.unistra.fr}
\affiliation{Institut de Physique et Chimie des Mat\'eriaux de Strasbourg and NIE, UMR 7504, Universit\'e de Strasbourg and CNRS, 23 rue du L\oe{}ss, BP43, 67034 Strasbourg Cedex 2, France}

\author{Xianglong Li}
\affiliation{Center for Integrated
Nanotechnologies, Los Alamos National Laboratory, Los Alamos, New Mexico 87545}
\altaffiliation{Current address: National Center for Nanoscience and Technology, Beijing, 100190 P. R. China}

\author{Han Htoon}
\affiliation{Center for Integrated
Nanotechnologies, Los Alamos National Laboratory, Los Alamos, New Mexico 87545}

\author{Louis E. Brus}
\affiliation{Department of Chemistry, Columbia University, New York, NY 10027}

\author{Stephen K. Doorn}
\affiliation{Center for Integrated
Nanotechnologies, Los Alamos National Laboratory, Los Alamos, New Mexico 87545}

\author{Tony F. Heinz}
\affiliation{Departments of Physics and Electrical Engineering, Columbia University, New York, NY 10027}

\date{\today}

\begin{abstract}
We report a comprehensive study of the two-phonon inter-valley (2D) Raman mode in graphene monolayers, motivated by recent reports of asymmetric 2D-mode lineshapes in freestanding graphene. For photon energies in the range $1.53~\rm eV - 2.71 ~\rm eV$, the 2D-mode Raman response of freestanding samples appears as bimodal, in stark contrast with the Lorentzian approximation that is commonly used for supported monolayers. The transition between the freestanding and supported cases is mimicked by electrostatically doping freestanding graphene at carrier densities above $2\times 10^{11}~\rm cm^{-2}$. This result quantitatively demonstrates that low levels of charging can obscure the intrinsically bimodal 2D-mode lineshape of monolayer graphene, which can be utilized as a signature of a quasi-neutral sample.
In pristine freestanding graphene, we observe a broadening of the 2D-mode feature with decreasing photon energy that cannot be rationalized using a simple one-dimensional model based on resonant \textit{inner} and \textit{outer} processes. This indicates that phonon wavevectors away from the high-symmetry lines of the Brillouin zone must contribute to the 2D-mode, so that a full two-dimensional calculation is required to properly describe multiphonon-resonant Raman processes.  


\end{abstract}

\maketitle
\paragraph{\textbf{Introduction}}
Raman scattering has played a central role in the advancement of graphene science~\cite{Ferrari07,Malard09,Ferrari06,Gupta06,Graf07}. This reflects the strong electron-phonon coupling at the center ($\bm \Gamma$ point) and edges ($\bf  K$, $\bf K'$ points) of graphene's two-dimensional Brillouin zone~\cite{Piscanec04}, as well as its simple low-energy band structure (Dirac cones).
Zone-center optical phonons ($q=0$) predominantly give rise to first-order modes (\textit{e.g.} the G-mode). Due to momentum conservation, finite momentum phonons do not participate in first-order processes. Such phonons are, however, responsible for second-order resonant processes~\cite{Reich00}, such as the defect-related $D$ mode and its symmetry-allowed two-phonon overtone (2D-mode, also known as G' or D$^*$), which involve near zone-edge optical phonons~\cite{Maultzsch04,Ferrari06,Gupta06,Graf07,Narula08,Basko08,Venezuela11}. 
Since these processes interweave the electron and phonon dispersions, they lead to dispersion in the Raman spectra with respect to the excitation photon energy~\cite{Reich00} and provide information on the electron and phonon bands~\cite{Malard07}.  In particular, the dependence of the  2D-mode to electronic structure, which endows with sensitivity the thickness and stacking order of few-layer graphene, is widely used as a diagnostic for sample characterization.~\cite{Ferrari06,Graf07,Gupta06}. 

\begin{figure*}[!ht]
\begin{center}
\includegraphics[scale=0.51]{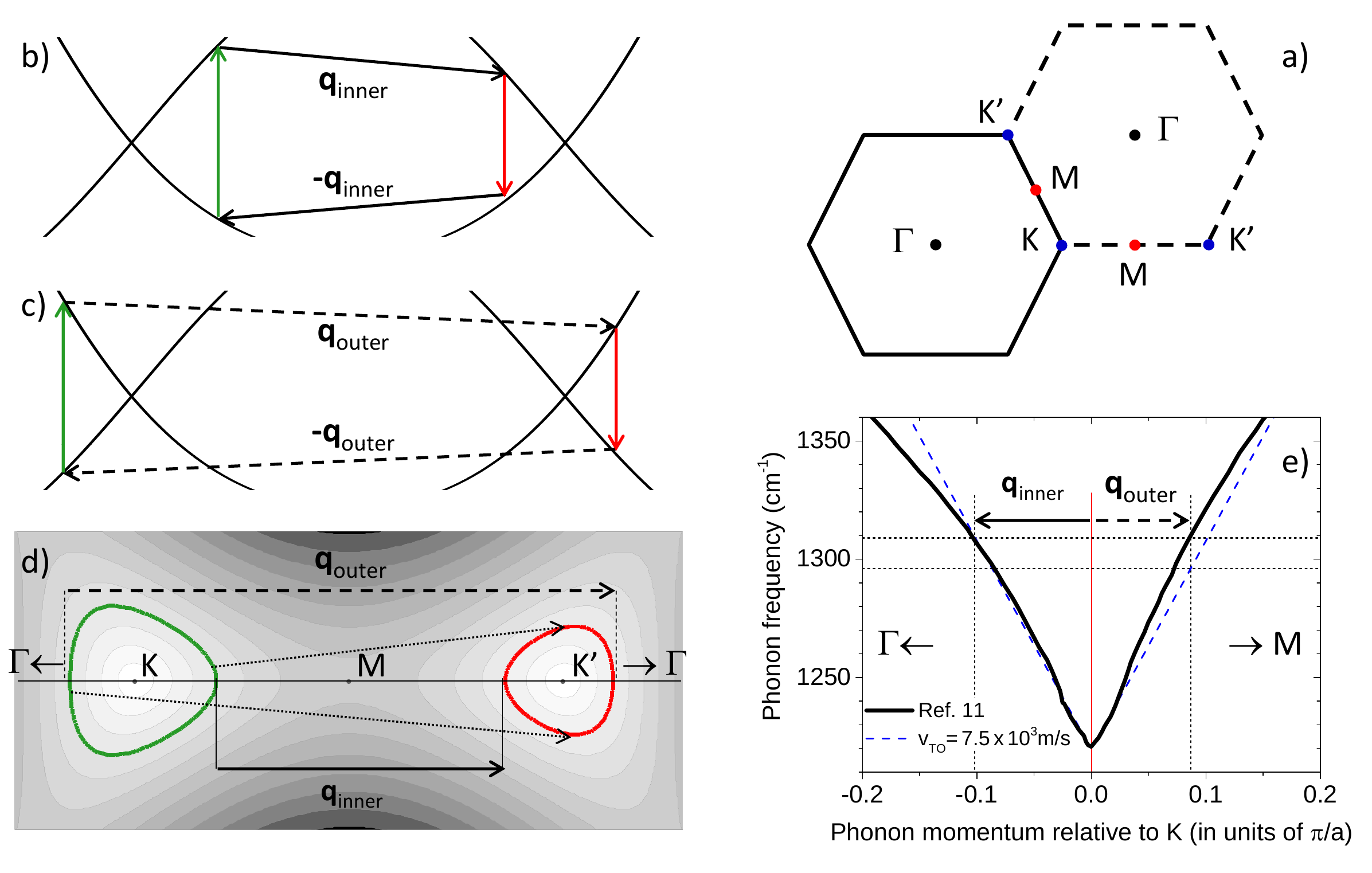}
\caption{(a) Brillouin zone of monolayer graphene (solid line), showing the high-symmetry lines. (b, c) One-dimensional, fully resonant, representation of the \textit{inner} (b) and \textit{outer} (c) loops in the momentum-energy space, along $\bm \Gamma-\bf K-\bf M-\bf K'-\bm \Gamma$. In this picture~\cite{Basko08}, an electron-hole pair is resonantly excited across the nearly-linear graphene bands by an incoming photon with energy $E_{L}$ (green vertical arrow). The electron and hole are resonantly scattered to a neighboring Dirac cone by a pair of optical phonons with energy $\hbar\omega_{\rm D}$ and opposite momenta. Finally, the electron-hole pair recombines radiatively and emits a photon $E_{L}-2\hbar\omega_{\rm D}$ (red arrow). (d) Schematic representation of the 2D-mode process in the two-dimensional Brillouin zone of graphene. The trigonal warping of the electronic dispersion -computed using a tight binding model~\cite{castronetoRMP2009}- is clearly visible. The green (red) lines represent the iso-energy contours at the incoming and outgoing photon energies. The solid (dashed) arrow corresponds to the \textit{inner} (\textit{outer}) loop. The dotted arrows illustrates contributions from optical phonons away from the high symmetry lines. (e) Theoretical dispersion of the TO phonon branch (black solid line, extracted from Venezuela \textit{et al.}~\cite{Venezuela11}) in the vicinity of the $\bm K$ point along the $\bm \Gamma- \bf K- \bf M$ line. The dashed blue line represents an isotropic phonon dispersion with $v_{\rm TO}=7.5\times 10^3 \rm ~m/s$. The solid (dashed) horizontal arrow represents the \textit{inner} (\textit{outer}) phonon momentum (relative to the $\bf K$ point) calculated for $E_L=1.5~\rm eV$, using the fully resonant one-dimensional models in a) (in  b)).}

\label{Sketch2D}
\end{center}
\end{figure*}

The 2D-mode is commonly described as a process that occurs chiefly along the high-symmetry lines ($\bf K-\bf M-\bm \Gamma-\bf K'-\bf M$)
of the graphene Brillouin zone~\cite{Maultzsch04,Ferrari06, Graf07}. This one-dimensional picture gives rise to an \textit{inner} and an \textit{outer} loop, involving phonon wavevectors $q<KK'$ ($q>KK'$), along the $\bf K \bm \Gamma$ (\textbf{KM}) direction (see \cref{Sketch2D}). Since both electron and phonon dispersions are anisotropic and display trigonal warping, we expect that the frequencies of the phonons selected by these two different channels will differ slightly~\cite{Maultzsch04}. Nevertheless, the 2D-mode feature of supported graphene monolayers is generally viewed as a single, quasi-Lorentzian peak~\cite{Ferrari06,Gupta06,Graf07} with a full width at half maximum (FWHM) $\Gamma_{\rm 2D}$ in the range $25-35~\rm cm^{-1}$, that is, on the same order of magnitude as that of the first order G-mode ($\Gamma_{\rm G}\sim14~\rm cm^{-1}$). This lineshape has been tentatively interpreted as a dominant contribution from the \textit{outer} loop~\cite{Kurti02,Ferrari06,Graf07}.

We have recently shown that freestanding graphene monolayers can be prepared that are not perturbed by doping or strain from the substrate. This conclusion was based on a spatially resolved study of the Raman G-mode feature. Under these conditions, we also observed that the 2D-mode feature displayed a narrower and bimodal lineshape~\cite{Berciaud09}, in stark contrast with the Lorentzian approximation that is commonly used for supported graphene monolayers. Similar lineshapes have been consistently observed in freestanding graphene~\cite{Luo12,Yoon12}, and the two subfeatures have been tentatively assigned by researchers to the \textit{inner} and \textit{outer} processes~\cite{Luo12}. Nevertheless, in spite of these phenomenological observations~\cite{Berciaud09, Luo12}, a rigorous explanation for the peculiar 2D-mode lineshape observed in pristine freestanding graphene monolayers, as well as for the striking differences between the lineshapes observed for freestanding versus supported samples, remains highly desirable.

In this letter, the 2D-mode lineshapes of both freestanding and supported graphene monolayers are thoroughly examined as a function of the photon energy, the polarization of the incoming and scattered photons, and the Fermi level. In freestanding samples, the 2D-mode appears, as previously reported~\cite{Berciaud09, Luo12}, as a bimodal feature, with an unanticipated energy-dependent lineshape. 
We find that electrostatic doping of freestanding graphene at densities above $2\times10^{11}~\rm cm^{-2}$ leads to a 2D-mode feature that is broader than for the undoped case and that appears to be nearly symmetric.   The lineshape observed for the 2D-mode for graphene samples supported on SiO$_2$ is also noticeably more symmetric than for the case of undoped freestanding graphene, but agrees well with the lineshape for the doped freestanding samples. This suggests that unintentional doping in such supported samples obscures the instrinic structure of the 2D-mode response. Although the general shape of the 2D-mode feature measured in pristine freestanding graphene can be rationalized in terms  of  \textit{inner}- and \textit{outer}-loop processes, we demonstrate that the evolution of the lineshape with photon energy is incompatible with a model based on a simple superposition of these channels. Our results thus show that phonons with wavevectors away from the high symmetry lines contribute to the 2D-mode. A full two-dimensional model is accordingly needed to properly describe the 2D-mode and, more generally, multi-phonon resonant processes, as has been proposed theoretically~\cite{Venezuela11,Narula12,MayPRB2013}.

\paragraph{\textbf{Experimental procedure and 2D-mode spectra}}
Graphene monolayers were deposited by mechanical exfolation on Si/SiO$_2$ substrates  pre-patterned with $4~\rm \mu m$ wide trenches, as described in Ref.~\cite{Berciaud09}. This process results in regions of graphene that were supported by the substrate, but also in freestanding regions.  In both cases, the graphene was not subject to any processing steps and is thus in its pristine state. The lack of doping of the freestanding portions of the graphene was established in a spatially resolved Raman study~\cite{Berciaud09,supp-info}.
Here, tunable Raman spectroscopy was performed under ambient conditions over the range $E_{L}=1.53~\rm eV-2.71~\rm eV$, using various laser sources (Ar-ion, HeNe, diode-pumped solid state and Ti:Sa). The optical radiation scattered by graphene was collected in a backscattering geometry and dispersed onto a charged-coupled device array by a single-pass optical spectrometer, with a spectral resolution better than $\sim 2 ~\rm cm^{-1}$.
The linearly polarized laser beam was focused onto a $\sim 1~\rm \mu m$ diameter spot. The laser power was maintained below $1~\rm mW$ to avoid laser-induced heating of the freestanding films and subsequent spectral shifts or lineshape changes.

\begin{figure*}[!htb]
\begin{center}
\includegraphics[scale=0.48]{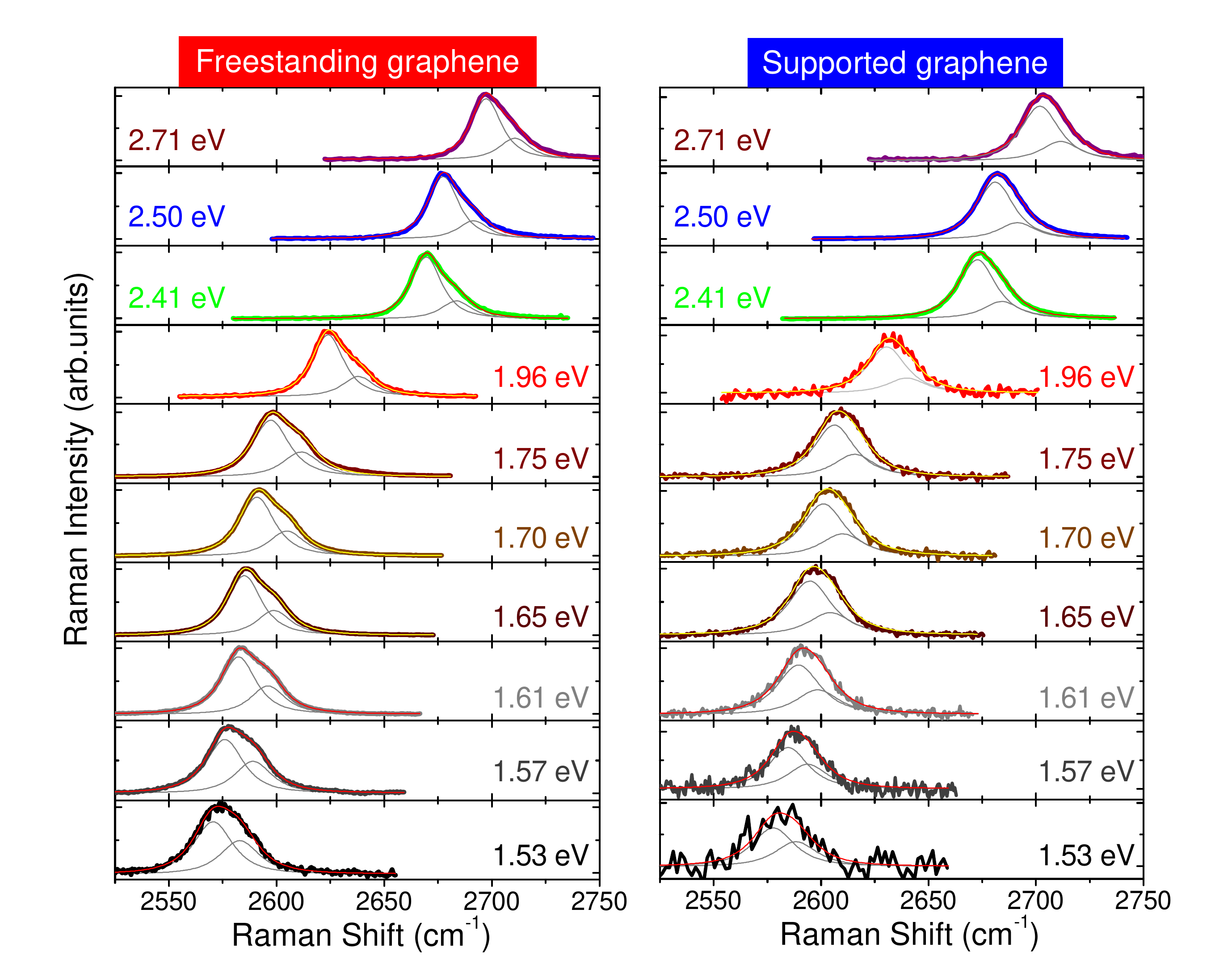}
\caption{Raman 2D-mode spectra  measured in the range $1.53~\rm eV - 2.71 ~\rm eV$ at the same spot on the freestanding (left panel) and supported  (right panel) areas of a same graphene monolayer. The thin colored lines are fits based on \cref{Baskonian}. The thin gray lines represent the 2D$^{\pm}$ subfeatures. The spectra measured on supported graphene were fit using the $\rm {I_{2D-}^{~}/I_{2D+}^{~}}$ integrated intensity ratio determined from the spectra measured on the freestanding part at the corresponding photon energy.}
\label{FigSpectra}
\end{center}
\end{figure*}

\begin{figure}[!ht]
\begin{center}
\includegraphics[scale=0.65]{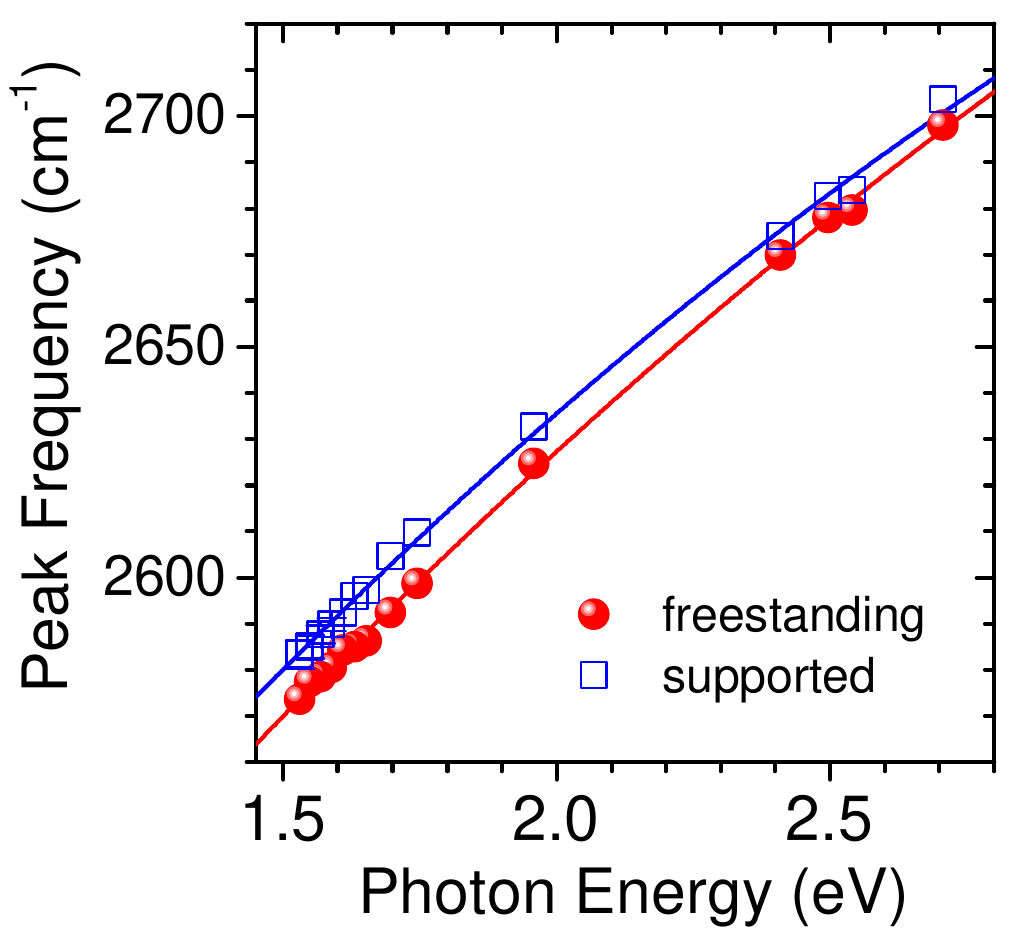}
\caption{Dispersion of the 2D-mode peak frequency measured on supported (squares) and freestanding (circles) graphene. The peak frequencies are extracted numerically. The solid lines are second-order polynomial fits.}
\label{FigDispersion}
\end{center}
\end{figure}

\cref{FigSpectra} shows a set of 2D-mode spectra measured on the same supported and freestanding regions of a graphene monolayer. From the study of the Raman G-mode, we estimate that the supported region of this sample exhibits a fairly low doping level of $n \sim 5\times10^{11} ~\rm cm^{-2}$~\cite{supp-info}. 
For both regions of the sample, we immediately see the expected the 2D-mode dispersion, \textit{i.e.} a Raman downshift with decreasing laser photon energies $E_{L}$ (see \cref{FigDispersion}). We also observe prominent energy-dependent lineshape changes on the freestanding part. 
In the visible range, the 2D-mode feature is fairly narrow (FWHM $\approx 22-24~ \rm cm^{-1}$), negatively skewed and shows negligible dependence on $E_{L}$. For $E_{L}<1.96\rm~eV$, two subfeatures, denoted $\rm 2D^{\pm}$ appear more prominently and we observe a modest broadening. In comparison, on the supported part, the 2D-mode lines are slightly broader, nearly symmetric and vary little with $E_{L}$ (see \cref{FigSpectra}). In \cref{FigDispersion}, we plot the numerically extracted peak position of the 2D-mode feature for all spectra. 
Interestingly, the 2D-mode feature appears consistently at lower frequency on the freestanding part than on the supported part of the sample, and its dispersion is slightly steeper on freestanding graphene~\cite{supp-info} (see~\cref{FigDispersion}). We also performed polarized Raman measurements on more than 10 freestanding samples at visible and near-infrared photon energies. We observed neither significant spectral shifts nor lineshape changes while rotating the sample and/or the analyzer with respect to the laser polarization~\cite{supp-info}.

\begin{figure*}[!htb]
\begin{center}
\includegraphics[scale=0.55]{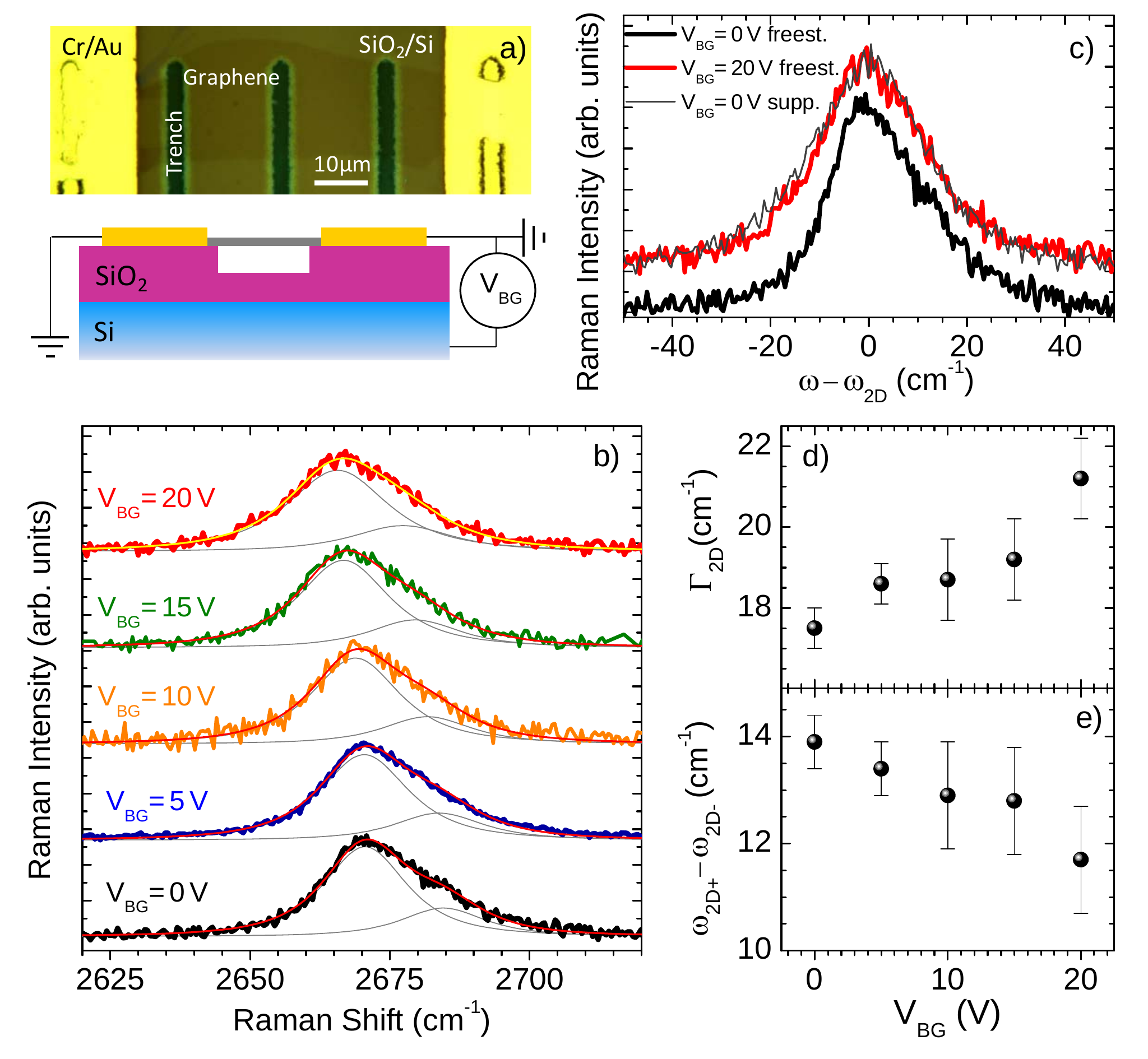}
\caption{(a) Sketch of a back-gated freestanding graphene device. b) 2D-mode spectra measured as a function of gate bias on freestanding graphene.  The spectra are vertically offset for clarity. The thin lines are fits based on \cref{Baskonian} using a fixed value of $\rm I_{2D-}/I_{2D+}=3.2$. c) Comparison between the 2D-mode lineshape of pristine freestanding graphene (black), electrostatically doped freestanding graphene (red) and supported graphene (thin gray line). (d) Linewidth of the $\rm  2D^{\pm}$ features and (e) Spectral shift $\omega_{\rm 2D+}-\omega_{\rm 2D-}$ extracted from the fits in b), as a function of gate bias. All spectra were measured at $E_{L}=2.33~\rm eV$ at a temperature of 6K.}
\label{FigLRT2Gating}
\end{center}
\end{figure*}

In order to investigate the influence of doping on the 2D-mode lineshape, we performed gate tunable measurements on the samples described above. Electrostatic gating was achieved by applying a bias between the Si substrate, used as a back gate, and a Ti/Au electrode contacting graphene, as shown in \cref{FigLRT2Gating}a. The Ti/Au electrode (1~nm/50~nm) was deposited by electron beam evaporation using a shadow mask. In the trenches, a residual $\rm SiO_2$ layer of 130~nm thickness was left in place to avoid gate leaks. Note that the devices were fabricated using a \textit{resist-free} process and were not contaminated by fabrication residues. In this study, the laser photon energy was $E_{L}=2.33~\rm eV$ and the power was kept below $200~\rm \mu W$.  The measurements were performed in vacuum ($\sim 10^{-5} \rm {mbar}$) and at low temperature (6K) in order to avoid gate leaks and obtain a stable electrostatic gating. 

\cref{FigLRT2Gating}b displays the low-temperature Raman 2D-mode spectra of the freestanding part of a graphene sample as a function of the gate bias. 
We estimate the charge density induced by gating of the freestanding graphene sample using a calculated value of $\rm 2.5\times 10^{10} cm^{-2}V^{-1}$ for the capacitance of the composite vacuum/oxide gate dielectric.  Assuming that graphene is neutral at zero bias, we then infer that the maximum voltage applied $(V_{\rm BG}=20~\rm V)$ corresponds to a Fermi energy of $\sim80~\rm meV$ relative to the Dirac point.
As the Fermi level is shifted, we observe a softening (by $3~\rm cm^{-1}$) and narrowing (by $4~\rm cm^{-1}$) of G-mode phonons~\cite{supp-info}.  
At the same time, we also observe clear changes in the 2D-mode lineshape (\cref{FigLRT2Gating}b). At zero bias, the 2D-mode displays the asymmetric lineshape that has been observed on all our freestanding, undoped samples at room temperature. At finite bias, the 2D-mode slightly softens by up to $\sim3~\rm cm^{-1}$ and becomes more symmetric. 
Interestingly, the lineshape observed on doped freestanding graphene is very similar to that of the neighboring supported part (\cref{FigLRT2Gating}c). 

\paragraph{\textbf{Phenomenological analysis}}
In order to rationalize the observation of bimodal lineshapes on freestanding graphene, we first assume that the 2D-mode is composed of two independent subfeatures $\rm 2D^{\pm}$, that arise from the \textit{inner} and \textit{outer} processes. Each component is described by an independent scattering intensity ($\rm I_{2D+}$ and $\rm I_{2D-}$) in fitting the experimental data based on the model proposed by Basko~\cite{Basko08}:
\begin{equation}
\frac{d\rm I_{\rm{2D^{\pm}}}\left(\omega\right)}{d\omega}\propto
\left[(\omega-\omega_{\rm 2D^{\pm}})^2+\frac{\Gamma_{\rm 2D}^2}{4\:(2^{2/3}-1)}\right]^{-\frac{3}{2}},
\label{Baskonian}
\end{equation}
where $\omega_{\rm 2D^{\pm}}$ and $\Gamma_{\rm 2D}$ denote the central frequency and FWHM of the $\rm 2D^{\pm}$ subfeatures, respectively. In this one-dimensional description, we assume a common value for the FWHM $\Gamma_{\rm 2D}$. The total intensities of the 2D$^\pm$ subfeatures (denoted $\rm I_{2D}^{\pm}$) are then readily obtained from an integration of \cref{Baskonian}.

\begin{figure*}[!tb]
\begin{center}
\includegraphics[scale=0.5]{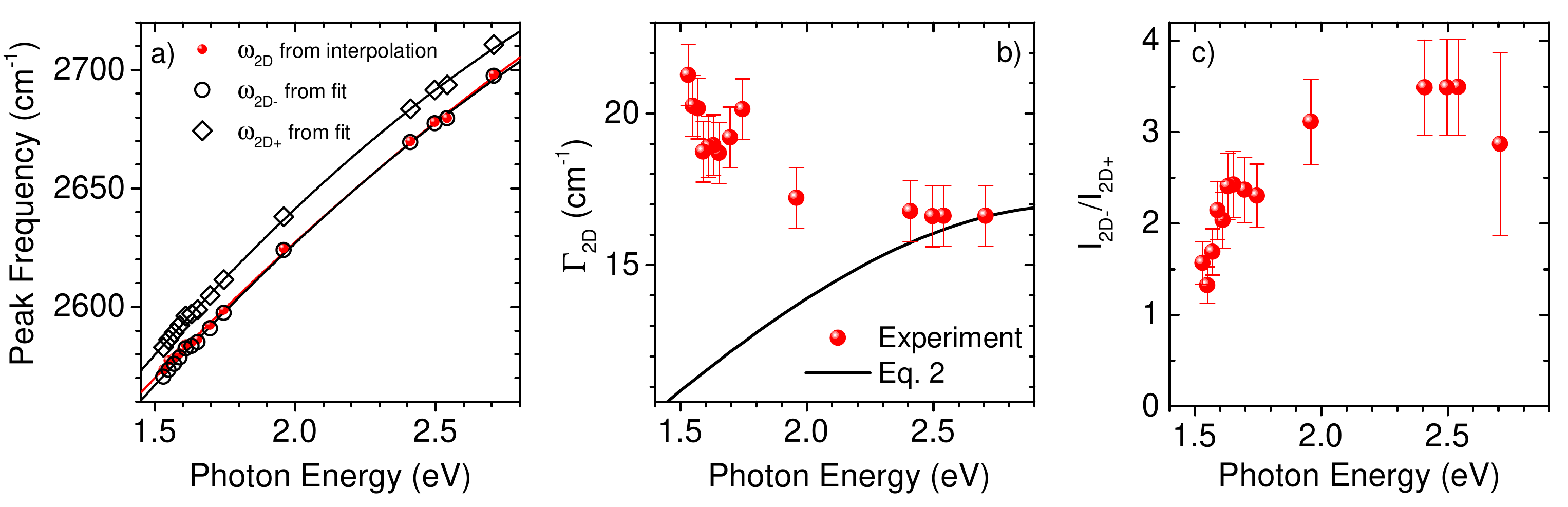}
\caption{(a) Dispersion of the 2D-mode peak frequency measured on freestanding (circles) graphene compared with that of the $\rm 2D^{\pm}_{}$ subfeatures (open symbols) obtained from a fit based on \cref{Baskonian}. The solid lines are second-order polynomial fits. (b) Full width at half maximum of the $\rm 2D^{\pm}_{}$ peaks and (c) Integrated intensity ratio of the $\rm 2D^-$ and $\rm 2D^+$ peaks as a function of the photon energy $E_{L}$. The solid line in b) is a theoretical estimation of $\Gamma_{\rm 2D}$ based on \cref{GammaBasko} considering the measured scaling of $v_{\rm TO}/v_{\rm F}$ with $E_L$ and theoretical calculations of the energy dependence of $\gamma_{eh}$~\cite{Venezuela11}. For clarity, the calculation has been normalized to match the value of $\Gamma_{\rm 2D}$ measured at $2.71~\rm eV$.}
\label{FigParamFits}
\end{center}
\end{figure*}

The 2D-mode spectra measured on supported samples and on electrostatically doped freestanding samples were also fit to \cref{Baskonian}. In these conditions, several sets of parameters may fit such nearly symmetrical lineshapes. To limit the number of adjustable parameters, we fit these spectra using the values of $ \rm {I_{2D-}^{~}/I_{2D+}^{~}}$ extracted from the fits of the 2D-mode spectra measured on pristine freestanting graphene (see \cref{FigParamFits} and \cref{FigLRT2Gating}c) at the corresponding photon energy. 

As shown in \cref{FigSpectra} and \cref{FigLRT2Gating}, fits based on \cref{Baskonian} reproduce our data very well. \cref{Baskonian} form gives rise to steeper spectral wings than a Lorentzian form, and better fits our data, with the same number of adjustable parameters~\cite{supp-info}. 
The resulting fitting parameters obtained on freestanding graphene are shown in \cref{FigParamFits}. 
The 2D$^{\pm}$ subfeatures have very similar dispersions that match that obtained from the numerically extracted peak  frequency of the composite 2D-mode feature (see \cref{FigParamFits}a). The splitting $\omega_{\rm 2D^+}-\omega_{\rm 2D^-}$ ranges from $\sim14~\rm cm^{-1}$ in the visible range down to $\sim12~\rm cm^{-1}$ at 1.53~\rm eV~\cite{supp-info}. 
The width $\Gamma_{\rm 2D}$ derived from the fits for freestanding graphene grows from $17~\rm cm^{-1}$ at 2.7~\rm eV up to $21~\rm cm^{-1}$ at 1.53~\rm eV (see \cref{FigParamFits}b). Another salient feature is the significant decrease of the $ \rm {I_{2D-}^{~}/I_{2D+}^{~}}$ ratio from $3.5$ in the visible range down to $1.5$ at $E_{L}=1.53~\rm eV$ as $E_{L}$ decreases (see \cref{FigParamFits}c).

\paragraph{\textbf{Discussion}}


As sketched in~\cref{Sketch2D}, the $\rm 2D$-mode involves a pair of near zone-edge transverse optical (TO) phonons (D-mode phonons~\cite{Maultzsch04}), with opposite momenta. In the energy range investigated here, the wavevectors of the D-mode phonons are significantly away from the Kohn anomaly at the $\bf K$ point (see \cref{Sketch2D}e)~\cite{Piscanec04}. As a result, contributions to the 2D-mode linewidth from the decay of optical phonons into resonant electron-hole pairs, which are known to be important for the G-mode phonon, can be neglected here for the D-mode phonons~\cite{Yan07,Das08}.
The  matrix element of the Raman 2D-mode  has to be evaluated using $4\rm ^{th}$ order perturbation theory, by summing all the scattering processes that satisfy energy and momentum conservation over the entire two-dimensional Brillouin zone of graphene~\cite{Maultzsch04,Narula08,Basko08,Venezuela11}. Thus, a broad range of electronic states and phonon wavevectors may participate, and produce a broadened 2D-mode feature, which, \textit{a priori}, is not expected to be a single symmetric peak, let alone have Lorentzian lineshape. 
This \textit{phonon broadening}, arising from trigonal warping effects may add to the broadening, induced by the finite lifetime of the electrons and holes involved in this Raman process, and much smaller contributions, on the order of 2~cm$^{-1}$, due to the anharmonic lifetime of the optical phonons~\cite{Bonini07,YanPRB2009,WuNL2012} and our experimental spectral resolution. Conversely, in the absence of any anisotropy in the electron and phonon dispersions, the \textit{inner} and \textit{outer} loops give rise to identical  2D$^\pm$ subfeatures, and the linewidth of the 2D-mode feature is expected to be given by~\cite{Basko08}:

\begin{equation}
\Gamma_{\rm 2D}=4\sqrt{2^{2/3}-1}\: \frac{v_{\rm{TO}}^{~}}{v_{\rm{F}}^{~}} \gamma_{eh},
\label{GammaBasko}
\end{equation}
where $v_{\rm{TO}}$ ($v_{\rm{F}}$) is the phonon (Fermi) velocity, defined as the slope of the phononic (electronic) dispersion at the phonon (electron) momentum corresponding to a given laser energy $E_L$, and $\gamma_{eh}$ is the electronic broadening parameter.

Let us first consider the differences between the spectra measured on pristine freestanding graphene and those on doped freestanding graphene (see \cref{FigLRT2Gating}d). Starting from the reference measured on pristine freestanding graphene, our fits imply a clear broadening as well as a modest reduction (by $\rm \sim ~2 cm^{-1}$) of the frequency difference between the $\rm 2D^{\pm}$ subfeatures (see \cref{FigLRT2Gating}e) as the doping level increases. The latter observation presumably results from small modifications of the electron and phonon dispersions upon doping~\cite{Maciel08}. We attribute the doping-induced broadening to an enhancement of $\gamma_{eh}$ in the presence of additional charge carriers~\cite{Basko09}.
Importantly, the fits of the 2D-mode spectra of pristine supported graphene (see \cref{FigSpectra}, right panel) reveal similar behaviors for the peak frequencies and more importantly for the width $\Gamma_{\rm 2D}$ ~\cite{supp-info}.

Our studies on gated freestanding and pristine supported samples thus demonstrate quantitatively that minor levels of doping ($\sim2\times10^{11}~\rm cm^{-2}$) and charge inhomogenity suffice to significantly alter the intrinsic Raman features of graphene. As a result, we are able to mimic the transition from a pristine freestaning graphene molonayer to an unintentionally doped supported one by simply tuning the charge carrier density. 
We conclude that the 2D-mode of supported graphene can, in first approximation, be regarded as a broadened copy of the intrinsic 2D-mode feature observed on pristine freestanding graphene.

Having established that the peculiar 2D-mode lineshapes observed on freestanding graphene are an indication of a quasi-undoped sample, we now analyze the intrinsic 2D-mode lineshape and first consider the scaling of $\Gamma_{\rm 2D}$ with $E_{L}$.
The electron and phonon dispersion of graphene show non-negligible deviations from linearity over the broad energy range probed here~\cite{Venezuela11}. Hence, the  measured dispersion of the 2D$^\pm$ subfeatures is not linear and is better fit to a second order polynomial cuve, the slope of which is twice the ratio of the TO phonon and Fermi velocities\cite{Mafra07}, $\frac{v_{\rm TO}^{}}{v_{\rm F}^{}}$, at the corresponding photon energy. 
The observation of a bimodal lineshape suggests that trigonal warping effects are not negligible. Therefore, one might \textit{a priori} consider different values of $\Gamma_{2D}$ for the 2D$^\pm$ subfeatures. However, as shown in \cref{FigParamFits}a, since the dispersion of the 2D$^\pm$ subfeatures have nearly equal slopes, the ratio $\frac{v_{\rm TO}^{}}{v_{\rm F}^{}}$ can be taken as identical for the \textit{inner} and \textit{outer} processes. In addition, at a given energy, $\gamma_{eh}$ can also be regarded as independent of the electron wavevector~\cite{Venezuela11}. This justifies our initial assumption of an identical linewidth $\Gamma_{\rm 2D}$ for the 2D$^\pm$ subfeatures. In a one-dimensional picture based on the \textit{inner} and \textit{outer} loops, $\Gamma_{\rm 2D}$ should follow the scaling predicted by \cref{GammaBasko}.

On freestanding graphene, the slope of the dispersion of the 2D$^\pm$ subfeatures increases from  $90~\rm cm^{-1}/\rm eV$ $(\frac{v_{\rm TO}^{}}{v_{\rm F}^{}}=5.5\times10^{-3})$ at $E_{L}=2.71~\rm eV$ up to  $120~\rm cm^{-1}/eV$ $(\frac{v_{\rm TO}^{}}{v_{\rm F}^{}}=7.5\times10^{-3})$ at $E_{L}=1.53~\rm eV$. Following \cref{GammaBasko}, this implies an increase of $\Gamma_{\rm 2D}$ by $\sim30\%$ as $E_{L}$ decreases, assuming that there is no change in the electronic broadening parameter $\gamma_{eh}$.
Thus, the $\sim20\%$ increase of $\Gamma_{\rm 2D}$ width decreasing $E_{L}$ agrees qualitatively with the increase of $\frac{v_{\rm TO}^{}}{v_{\rm F}^{}}$. However, this broadening is expected to be largely overcome by the decrease with energy in $\gamma_{eh}$.
Indeed, in the low-energy range, where the graphene bands are quasi-linear, the measured electronic linewidth scales roughly linearly with $E_{L}$~\cite{Li09b,Knox11}. From \cref{GammaBasko}, considering the partial compensation of the decrease of $\gamma_{eh}$ by the increase of $\frac{v_{\rm TO}^{}}{v_{\rm F}^{}}$, we should expect a narrowing of $\Gamma_{\rm 2D}$ from $\sim 17 \rm cm^{-1}$ to $\sim 11 \rm cm^{-1}$ as $E_{L}$ decreases from $2.71~\rm eV$ to  $1.53~\rm eV$~\cite{Venezuela11}.
 As a result, the scaling of $\Gamma_{\rm 2D}$ derived from \cref{GammaBasko} is grossly inconsistent with our experimental data (see \cref{FigParamFits}b).We conclude that the observed broadening of the 2D-mode feature cannot be rationalized if one considers only the \textit{inner} and \textit{outer} processes. This suggests that a full two-dimensional description of the 2D-mode (see \cref{Sketch2D}c) is more appropriate.

Recent \textit{ab initio} calculations of multi-phonon resonant Raman modes in unstrained graphene using such two-dimensional models have suggested that the contours of the phonon wavevectors that contribute to the 2D-mode closely  follow the iso-energy contours of the phonon dispersion~\cite{Venezuela11, Narula12}. In other words, electron and phonon trigonal warping effects are opposite~\cite{Gruneis09}, and partially cancel out.  In particular, over the energy range investigated here, the \textit{calculated} frequency shift between \textit{inner} and \textit{outer} phonons is nearly constant and only on the order of a few $\rm cm^{-1}$~\cite{Venezuela11}. An illustration is shown in \cref{Sketch2D}e. For $E_L=1.5~\rm eV$, a difference of $\sim 1~\rm cm^{-1}$ only is predicted for the frequencies of the \textit{inner} and \textit{outer} D-mode phonons.
 As a consequence, a quasi-symmetric feature, with a linewidth that essentially follows the scaling proposed by Basko (see \cref{GammaBasko}), is expected theoretically.
In our study, however, the observation of two $\rm 2D^{\pm}$ subfeatures separated by a nearly constant shift of $\sim12-14 ~\rm cm^{-1}$ suggests that the compensation of trigonal warpings of the electronic and phononic dispersion relations is imperfect. This allows for the observation of a broader range of phonon energies, such that $\Gamma_{\rm 2D}$ is not simply proportional to $\gamma_{eh}$ (see \cref{GammaBasko}). We would like to stress that our experimental results cannot be explained without a partial compensation of the trigonal warping effects. Indeed, considering electronic trigonal warping alone~\cite{Venezuela11}, and assuming an isotropic phononic dispersion relation, we estimate, using our measured value of $v_{\rm TO}/v_{\rm F}=7.5\times10^{-3}$ at $E_L=1.5~\rm eV$ with $v_{\rm F}\sim 10^6~\rm m s^{-1}$~\cite{castronetoRMP2009}, that the splitting between the $\rm 2D^{\pm}$ subfeatures would be as large as $\sim \rm 26~cm^{-1}$ at $E_{L}=1.5~\rm eV$ (see \cref{Sketch2D}e), and would grow with increasing $E_L$. This is in contradiction with the much smaller and nearly energy-independent values of $12-14~\rm cm^{-1}$ measured here.

In principle, polarization resolved measurements could help unravel the contributions from 2D-mode phonons with varying angle along and away from the high-symmetry lines of the graphene Brillouin zone. Indeed, as demonstrated by Narula \textit{et al.} the calculated footprint of the dominant phonon wavevectors vary with the polarization configuration~\cite{Narula12}. However, due to the hexagonal symmetry of the graphene Brillouin zone, and the partial compensation of electronic and phononic trigonal warpings, the contributing phonons may produce a virtually polarization independent lineshape in the near-IR and visible range, as we observed experimentally~\cite{supp-info}. 
In any event, calculations by Venezuela \textit{et al.} imply that, contrary to previous predictions based solely on the phonon density of states~\cite{Kurti02, Ferrari06, Luo12}, matrix-element effects cause phonons near the \textit{inner} wavevector to provide most of the 2D-mode intensity~\cite{Venezuela11}. A similar conclusion was reached by several groups who studied the behavior of the 2D-mode feature under uniaxial strain~\cite{Huang10,Yoon11,Frank11} and the dispersion of the 2D-mode feature in supported graphene bilayers~\cite{Mafra11}.
Consequently, the prominent $\rm 2D^{-}$ feature observed in the visible range is tentatively assigned to phonon wavevectors in the vicinity of the $\bf K- \bm \Gamma$ direction (\textit{i.e.}, near the \textit{inner} loop).

Finally, our observation of slightly broader $\rm 2D^{\pm}$ subfeatures with a ratio of $\rm {I_{2D-}^{~}/I_{2D+}^{~}}$ decreasing from $3.5$ in the visible range down to $1.5$ at $E_{L}=1.53~\rm eV$ (see \cref{FigParamFits}) is in qualitative agreement with the calculated angular distribution of the 2D-mode intensity, which is expected to become more isotropic as $E_{L}$ decreases (compare Figures 25 and 26 in Ref.~\cite{Venezuela11}). 
This supports the claims that electron and phonon trigonal warpings do not fully compensate one another and that phonon modes away from the high-symmetry lines contribute significantly to the broadening of the 2D Raman feature. 
We note, however, that in the low-energy limit, trigonal warping effects vanish and the 2D$^{\pm}$ subfeatures are thus expected to merge into one symmetric feature. Experiments outside the range of photon energies investigated here, as well as more detailed \textit{ab initio} calculations of the 2D-mode lineshape, should provide further insights into the contributions of phonons away from the high symmetry lines to multiphonon Raman processes.

\paragraph{\textbf{Conclusion}}

For pristine freestanding graphene, we have found that the 2D-mode feature exhibits a markedly asymmetric lineshape for measurements with laser excitation energies between $1.53~\rm eV$ and $2.71 ~\rm eV$.  The experimental profile can be fit accurately using the sum two features of a modified Lorentzian profile predicted theoretically for the electronically resonant 2D-mode process with short-lived electronic excitations. The existence of two defined spectral components is particularly apparent at lower excitation photon energy.  The spectral shift between these two components is on the order of $12-14~\rm cm^{-1}$ and shows a relatively weak dependence on laser excitation energy.  
Our experiments show that the observed asymmetry and bimodal character of the 2D-mode spectra become very slight when relatively low levels of charge ($\sim 2\times 10^{11}~\rm cm^{-2}$) are injected into the freestanding graphene monolayer.  Within our phenomenological treatment of the lineshape, we can account for this change by enhanced broadening of the two components, as would be expected for an increased electronic relaxation rate.  Similarly, for graphene samples supported by the typical SiO$_2$ substrate, we see only very weak asymmetry in the 2D-mode profile.  This change in lineshape compared to the pristine freestanding graphene sample can thus be naturally explained as a consequence of unintentional charging effects.  The present studies reveal the sensitivity of the 2D-mode to relatively small changes in the environment.
In addition, while the intrinsic 2D-mode linshape is compatible with a simple scheme of dominant contributions from phonons in high-symmetry regions of the \textit{inner} and \textit{outer} loops, the scaling of the widths of the two components with photon energy diverges significantly from the expected theoretical trends.   We conclude, in keeping with recent theoretical studies, that phonons away from these high-symmetry regions contribute significantly to the Raman 2D scattering process. Consequently, an accurate theory must be formulated within a full two-dimensional picture of the resonant Raman process in order to account the evolution of the 2D-mode feature with experimental conditions. More generally, these results provide an impetus for further theoretical studies that could result in a detailed explanation for the bimodal 2D-mode lineshape and its evolution with experimental conditions.
 
We are grateful to J. Maultzsch, F. Mauri, and R. Narula for inspiring discussions, and to M.Y. Han and M. Romeo for experimental help.
We acknowledge support from the MURI program AFOSR through grant FA9550-09-1-0705 and from the DOE through grant DE-FG02-11ER16224 for research at Columbia University, from the Universit\'e de Strasbourg, the CNRS and C'Nano GE for research carried out in France, and from the LANL LDRD program.  This work was performed in part at the Center for Integrated Nanotechnologies, a U.S. Department of Energy, Office of Basic Energy Sciences user facility.


\begin{thebibliography}{10}

\bibitem{Ferrari07}
A.~C. Ferrari, Solid State Communications {\bf 143},  47   (2007).

\bibitem{Malard09}
L. Malard, M. Pimenta, G. Dresselhaus, and M. Dresselhaus, Physics Reports {\bf
  473},  51   (2009).

\bibitem{Ferrari06}
A.~C. Ferrari {\it et~al.}, Physical Review Letters {\bf 97},  187401  (2006).

\bibitem{Gupta06}
A. Gupta {\it et~al.}, Nano Letters {\bf 6},  2667  (2006).

\bibitem{Graf07}
D. Graf {\it et~al.}, Nano Letters {\bf 7},  238  (2007).

\bibitem{Piscanec04}
S. Piscanec {\it et~al.}, Phys. Rev. Lett. {\bf 93},  185503  (2004).

\bibitem{Reich00}
C. Thomsen and S. Reich, Phys. Rev. Lett. {\bf 85},  5214  (2000).

\bibitem{Maultzsch04}
J. Maultzsch, S. Reich, and C. Thomsen, Phys. Rev. B {\bf 70},  155403  (2004).

\bibitem{Narula08}
R. Narula and S. Reich, Phys. Rev. B {\bf 78},  165422  (2008).

\bibitem{Basko08}
D.~M. Basko, Phys. Rev. B {\bf 78},  125418  (2008).

\bibitem{Venezuela11}
P. Venezuela, M. Lazzeri, and F. Mauri, Phys. Rev. B {\bf 84},  035433  (2011).

\bibitem{Malard07}
L.~M. Malard {\it et~al.}, Phys. Rev. B {\bf 76},  201401  (2007).

\bibitem{Kurti02}
J. K\"urti, V. Z\'olyomi, A. Gr\"uneis, and H. Kuzmany, Phys. Rev. B {\bf 65},
  165433  (2002).

\bibitem{Berciaud09}
S. Berciaud, S. Ryu, L.~E. Brus, and T.~F. Heinz, Nano Letters {\bf 9},  346
  (2009).

\bibitem{Luo12}
Z. Luo {\it et~al.}, Applied Physics Letters {\bf 100},  243107  (2012).

\bibitem{Yoon12}
J.-U. Lee, D. Yoon, and H. Cheong, Nano Letters {\bf 12},  4444  (2012).

\bibitem{castronetoRMP2009}
A.~H. Castro~Neto {\it et~al.}, Reviews of Modern Physics {\bf 81},  109
  (2009).

\bibitem{Narula12}
R. Narula, N. Bonini, N. Marzari, and S. Reich, Phys. Rev. B {\bf 85},  115451
  (2012).

\bibitem{MayPRB2013}
P. May {\it et~al.}, Phys. Rev. B {\bf 87},  075402  (2013).

\bibitem{supp-info}
See Supporting Information  .

\bibitem{Yan07}
J. Yan, Y. Zhang, P. Kim, and A. Pinczuk, Phys. Rev. Lett. {\bf 98},  166802
  (2007).

\bibitem{Das08}
A. Das {\it et~al.}, Nature Nanotechnology {\bf 3},  210  (2008).

\bibitem{Bonini07}
N. Bonini, M. Lazzeri, N. Marzari, and F. Mauri, Phys. Rev. Lett. {\bf 99},
  176802  (2007).

\bibitem{YanPRB2009}
H. Yan {\it et~al.}, Phys. Rev. B {\bf 80},  121403  (2009).

\bibitem{WuNL2012}
S. Wu {\it et~al.}, Nano Letters {\bf 12},  5495  (2012).

\bibitem{Maciel08}
I.~O. Maciel {\it et~al.}, Nat Mater {\bf 7},  1476  (2008).

\bibitem{Basko09}
D.~M. Basko, S. Piscanec, and A.~C. Ferrari, Phys. Rev. B {\bf 80},  165413
  (2009).

\bibitem{Mafra07}
D.~L. Mafra {\it et~al.}, Phys. Rev. B {\bf 76},  233407  (2007).

\bibitem{Li09b}
G. Li, A. Luican, and E.~Y. Andrei, Phys. Rev. Lett. {\bf 102},  176804
  (2009).

\bibitem{Knox11}
K.~R. Knox {\it et~al.}, Phys. Rev. B {\bf 84},  115401  (2011).

\bibitem{Gruneis09}
A. Gr\"uneis {\it et~al.}, Phys. Rev. B {\bf 80},  085423  (2009).

\bibitem{Huang10}
M. Huang, H. Yan, T. Heinz, and J. Hone, Nano Letters {\bf 10},  4074  (2010).

\bibitem{Yoon11}
D. Yoon, Y.-W. Son, and H. Cheong, Phys. Rev. Lett. {\bf 106},  155502  (2011).

\bibitem{Frank11}
O. Frank {\it et~al.}, ACS Nano {\bf 5},  2231  (2011).

\bibitem{Mafra11}
D. Mafra {\it et~al.}, Carbon {\bf 49},  1511   (2011).

\end{thebibliography}

\end{document}